\documentclass[%
; preprint,
twocolumn,
; a4paper,
superscriptaddress,
preprintnumbers,
showkeys,
showpacs,
amsmath,
amssymb,
pre,
aps]
{revtex4-1}

\usepackage{amsmath}    
\usepackage{amssymb}    
\usepackage{xfrac}

\usepackage{graphicx}   
\usepackage{textcomp}   

\begin{document}

\newcommand{\unit}[1]{\; \mathrm{#1}}
\newcommand{\e}[1]{\cdot 10^{#1}}
\newcommand{\lS}{$l_S$}
\def\smallfrac#1#2{{\textstyle\frac{#1}{#2}}}


\title{Coalescence of Islands in Freely-Suspended Smectic Films}

\author{Z.~H.~Nguyen}
\affiliation{Department of Physics and Soft Materials Research Center, University of Colorado, Boulder, CO, 80309, USA}
\affiliation{Center for High Technology Development, Hanoi, Vietnam}

\author{K.~Harth}
\affiliation{Institute for Physics, Otto von Guericke University Magdeburg,  D-39016 Magdeburg, Germany}
\affiliation{Max Planck Center for Complex Fluid Dynamics, University of Twente, 7500AE Enschede, The Netherlands}

\author{A.~M.~Goldfain}
\affiliation{Department of Physics and Soft Materials Research Center, University of Colorado, Boulder, CO, 80309, USA}

\author{C.~S.~Park}
\affiliation{Department of Physics and Soft Materials Research Center, University of Colorado, Boulder, CO, 80309, USA}

\author{J.~E.~Maclennan}
\affiliation{Department of Physics and Soft Materials Research Center, University of Colorado, Boulder, CO, 80309, USA}

\author{M.~A.~Glaser}
\affiliation{Department of Physics and Soft Materials Research Center, University of Colorado, Boulder, CO, 80309, USA}

\author{N.~A.~Clark}
\affiliation{Department of Physics and Soft Materials Research Center, University of Colorado, Boulder, CO, 80309, USA}

\begin{abstract}
Smectic liquid crystal films a few molecular layers thick that are freely suspended in air are used as a model system to study the coalescence of fluids in two dimensions. High-speed video microscopy is used to observe the coalescence of islands, which are thicker, disk-shaped regions of the film, in a process driven by the line tension associated with edge dislocations along the island boundaries and limited by viscous dissipation in the liquid crystal and in the surrounding air.
The early time growth of the bridge connecting the merging islands reveals much slower dynamics than predicted by Hopper's classical hydrodynamic model of coalescence of two infinitely long, fluid cylinders in vacuum, a discrepancy proposed to be due to significant dissipation in the background film and in the air that is not included in Hopper's theory. At late times, the elliptical merged island relaxes exponentially to a circular shape, at rates that are described quantitatively by a model originally developed for the evolution of fluid  domains in Langmuir films.
\end{abstract}


\maketitle

\section{Introduction}

The coalescence of two fluid objects and the converse process, the breakup of fluid drops, are simple, beautiful examples of singular physical phenomena involving the divergence of a physical quantity. Early work in this field was carried out by J.~J.~Thomson, who studied the evolution of ink droplets as they merged with water \cite{thomson_formation_1885}. More recent topics of investigation include the dynamics of drop pinch-off \cite{shi_cascade_1994,doshi_persistence_2003} and the expansion of the fluid bridge connecting two coalescing drops \cite{aarts_hydrodynamics_2005,case_coalescence_2008,Sprittles2012,Xia2019,Klopp2020}. Hopper's hydrodynamic modeling of the coalescence dynamics of two infinite, fluid cylinders \cite{hopper_coalescence_exact_1984,hopper_coalescence_1993} and further works by Eggers, Lister and Stone on the crossover between viscous and inertial regimes during coalescence \cite{eggers_coalescence_1999} have stimulated several experimental and theoretical studies of both three-dimensional ($3$D) \cite{menchaca-rocha_coalescence_2001,burton_role_2007,paulsen_viscous_2011,hack_selfsimilar_2020,Anthony2020} and quasi-two-dimensional ($2$D) fluids \cite{mann_hydrodynamics_1995,delabre_coalescence_2010,Nguyen2011,Petrov2012,Shuravin2019}. Fluid membranes represent near-ideal model systems for studying hydrodynamics in reduced dimensions, described theoretically by Saffman and Delbr\"uck \cite{SaffmanDelbruck1975,Saffman1976}. Usually such membranes have finite thickness and are immersed in another fluid (or are confined to a fluid-fluid interface) and so are only quasi-two-dimensional. Nevertheless, they can display $2$D character when the Saffman length, the ratio between the $2$D and $3$D viscosities, is sufficiently large, as we have demonstrated experimentally for inclusions in smectic films \cite{nguyen_crossover_2010, eremin_two-dimensional_2011,Qi2014,Kuriabova2016,Qi2016}. Saffman and Delbr\"uck's pioneering treatment of hydrodynamics in $2$D  has been extended recently to describe $2$D hydrodynamic interactions between embedded particles \cite{Oppenheimer2011,DiLeonardo2011} and the shape relaxation of fluid lipid domains \cite{Camley2010}.

Fluid smectic liquid crystal (LC) films freely suspended in air are near-ideal realizations of homogeneous $2$D fluids that can be used as model $2$D hydrodynamic and structural systems. We describe experiments using high-speed video  microscopy to observe the coalescence of pairs of flat, disk-shaped fluid smectic islands embedded in thinner background films of the same LC material, in events driven by the line tension along the island boundaries. In these investigations, pairs of islands with similar radii and the same thickness merge to form single islands of the same total area.  The coalescence dynamics were characterized by measuring the width of the bridge connecting the merging islands as a function of time.

Analysis of such events is based on treatment of the islands as incompressible, viscous discs of  fixed thickness much smaller than their radius, embedded in the background film and in Saffman-Delbr\"uck contact with the surrounding air.  A quantitative hydrodynamic description of the dynamics of coalescence in such a system is available only for later times in the relaxation, where a model by Mann et al.\ \cite{mann_hydrodynamics_1995} predicts that the merged island shape becomes elliptical and relaxes exponentially to a circular shape in a time that depends on the island boundary line tension, the LC and air shear viscosities, and the island  and background film  thicknesses.  Measurements of the relaxation times in the smectic~A phase of the liquid crystal $8$CB, where these materials properties are known, enables testing and verification of the Mann model using no fitting parameters,  confirming that accounting for dissipation in the island, air, and the background film is necessary and sufficient to describe quantitatively the relaxation at long times.

The shape relaxation of a thin, quasi-$2$D fluid domain of equilibrium radius $R$ is predicted in Mann's theory \cite{mann_hydrodynamics_1995} to occur in a characteristic time that depends on its reduced radius $\varepsilon = R/\ell_S$, where, in the case of a smectic film that is in contact with the air both above and below, the Saffman length is given  by $\ell_S=\eta_{2D}/(2\eta')$. When $\varepsilon \ll 1$, most of the dissipation occurs in the film and the relaxation has two-dimensional character \cite{mann_hydrodynamics_1995}, while when $\varepsilon \gg 1$, the dominant dissipation is in the surrounding air and the dynamics are three-dimensional. In previous experiments on the shape fluctuations of fluid domains in Langmuir films, typically $\varepsilon = {\cal O}(100)$ \cite{stone_hydrodynamics_1995}. However, in our freely-suspended smectic films, $\varepsilon$ can be tuned in the critical range from $0.1$ to $1$ by varying the film thickness in order to control $\ell_S$ \cite{nguyen_crossover_2010}. As we shall see below, in this regime small deviations from the equilibrium domain shape decay at a rate dictated  by dissipation in both $2$D and $3$D, the overall relaxation time constant $\tau_r$ being the sum of the characteristic times associated with these individual modes, i.e., $\tau_r=\tau_{2D}+\tau_{3D}$ \cite{mann_hydrodynamics_1995}.

At early times, it is instructive to apply Hopper's model describing the coalescence of two fluid cylinders in a hypothetical inviscid fluid  (summarized in Appendix~A) to the island coalescence case, under the assumption that the effects of flow of the surrounding air and of the background film may be ignored.  While we find that the observed functional form of the bridge width dynamics  is similar to the Hopper model, in agreement with the results of Shuravin et al.\  \cite{Shuravin2019},  our experiments yield measured characteristic growth rates that are only about half the Hopper prediction, contradicting Shuravin's claim of overall agreement with Hopper's model.  This indicates that the dissipative effects of both the surrounding air and the background film must also be included at short times.

\begin{figure}
 \centering
 \includegraphics[width=3in]{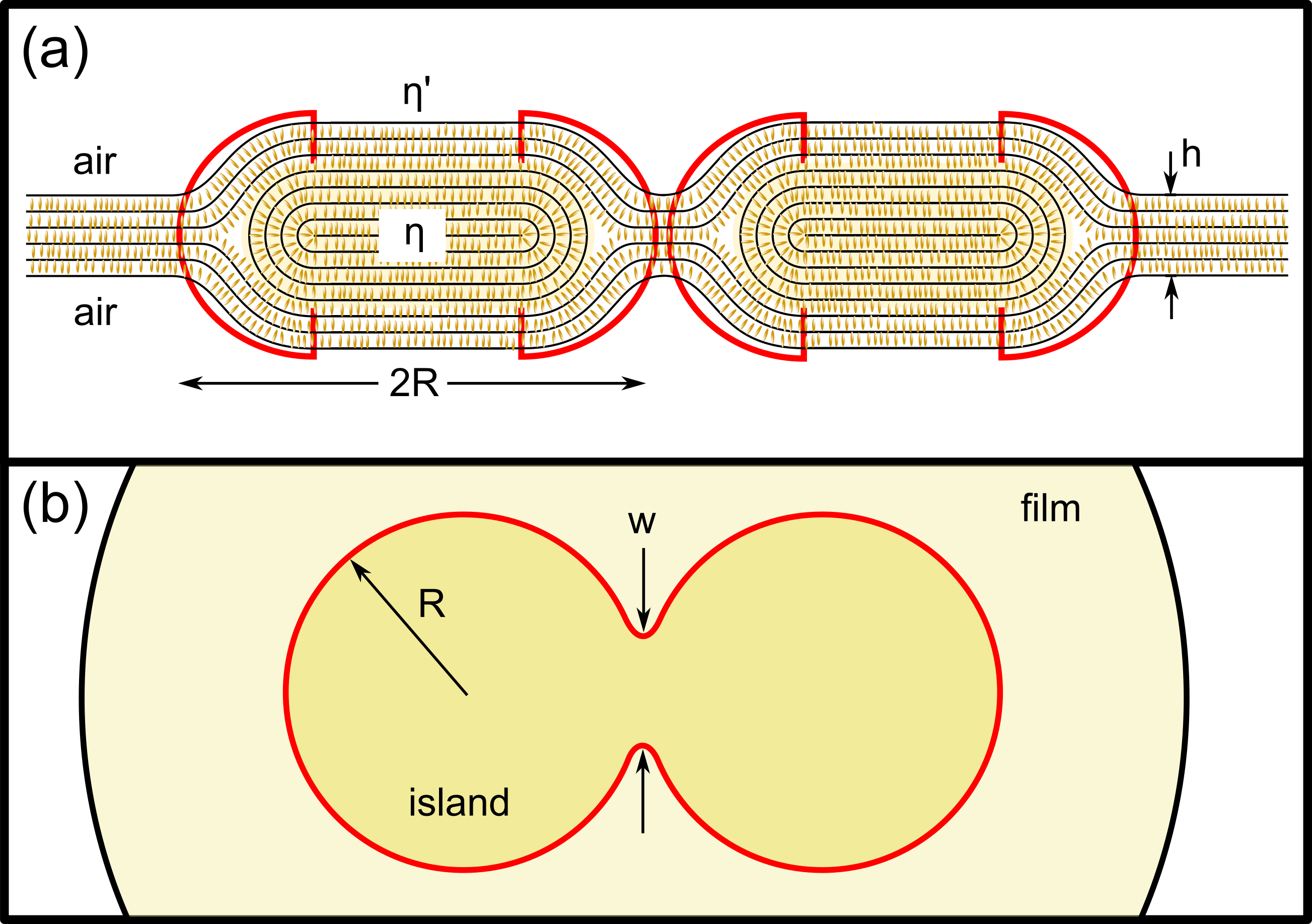}
 \caption{Coalescing islands in a smectic fluid film. (a) Cross-section of two merging $14$-layer islands in a $5$-layer background film. The merged island is bounded by an edge-dislocation loop (with the cross-section indicated by red lines). (b) Top view of island coalescence showing a bridge of width $w$. Before the islands make contact, they are kept circular by the line tension associated with the edge dislocations at their boundaries. This line tension drives the expansion of the bridge connecting the two islands when they coalesce, as well as the relaxation of the merged island to a circular shape at later times.}
         \label{fig:Sketch}
        \end{figure}

\section{Islands in thin smectic films}

Smectic liquid crystals can be drawn into stable films freely-suspended in air.  Because of the smectic layering, such films are of uniform thickness and are homogeneous in their materials properties such as viscosity \cite{young_light-scattering_1978}. Smectic films have low vapor-pressure and exhibit Newtonian viscosity, making them useful for hydrodynamics experiments in $2$D.
Islands can appear spontaneously in the films or can be created intentionally using airjets \cite{pattanaporkratana_manipulation_2004}.
These islands are bounded by edge-dislocation loops, sketched in Fig.~\ref{fig:Sketch}, with
an energy per unit length that may be expressed as a line tension $\lambda \propto \Delta N=N_i-N_b$, where $N_i$ and $N_b$ are respectively the number of molecular layers in the island and background films \cite{geminard_meniscus_1997}. This line tension minimizes the island perimeter and acts as the driving force in island coalescence. The dynamics are limited by dissipation arising from the bulk viscosity of air $\eta'$ and the $2$D viscosity of the film $\eta_{2D}=\eta h$, where $\eta$ is the $3$D viscosity of the liquid crystal and $h=Nd$ is the island or background film thickness, with $d$ being the thickness of a single smectic layer. The absence of a substrate in these experiments allows us to obtain high-contrast images of the islands using reflection-mode microscopy. The film and island layer numbers were determined by optical reflectivity \cite{rosenblatt_reflectivity_1980,pankratz_optical_reflectivity_2000}.

\begin{figure}[hbt]
 \centering
 \includegraphics[width=3in]{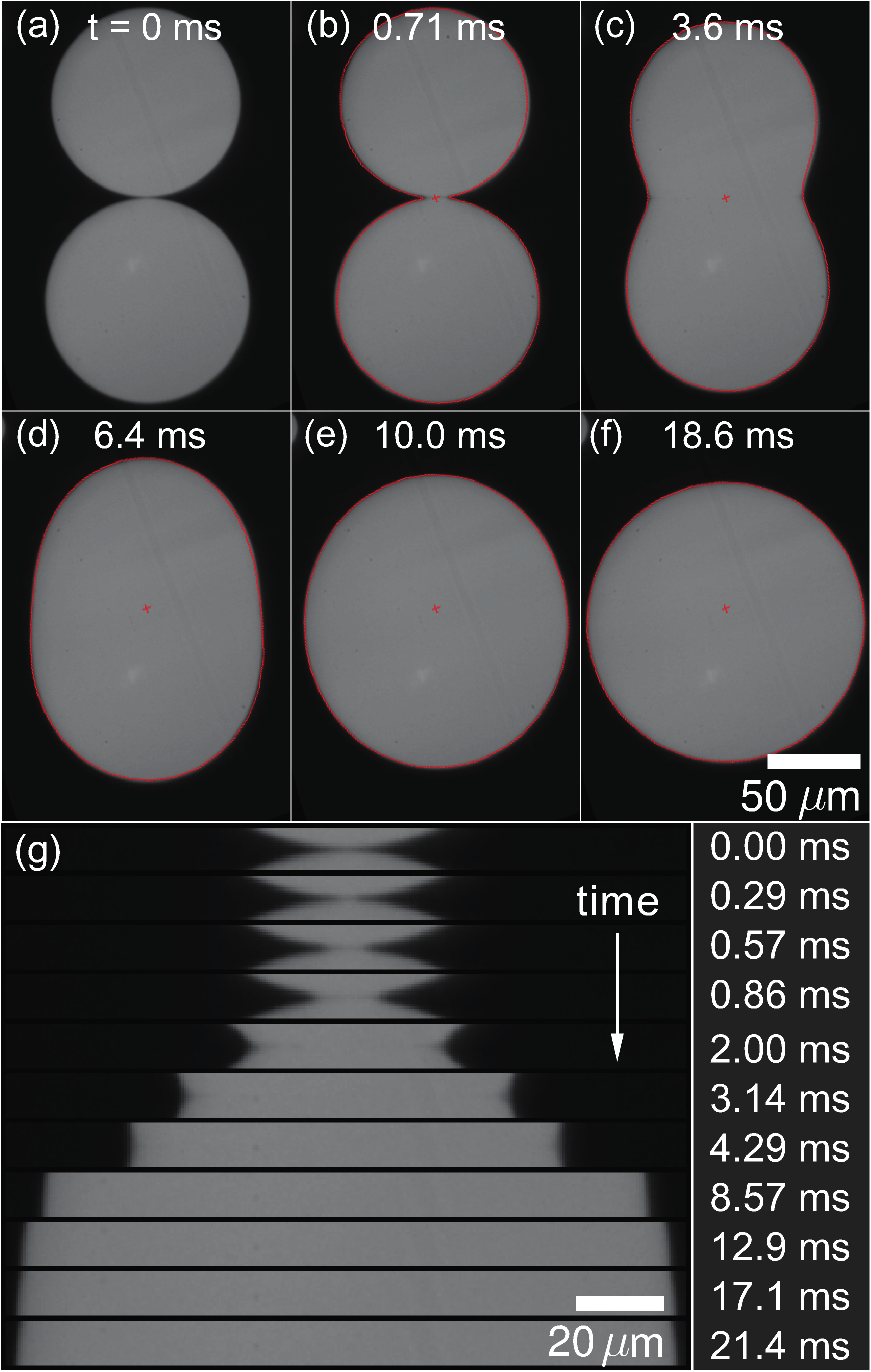}
 \caption{Coalescence of two islands in a fluid smectic film. (a) Typical coalescence event (starting at $t=0$) of two disk-like islands (bright, circular domains) in a thin smectic~A film (dark surroundings). The background film is four layers thick ($N_b=4$), and both islands have $N_i=27$. The original island radii are $50.7 \unit{\mu m}$ and $54.6 \unit{\mu m}$. The bridge connecting the merging islands expands rapidly at first (b,c), while at later times (d--f) the merged island relaxes slowly to its final circular shape. The red outlines are Fourier fits to the boundary shape obtained using Eq.~\ref{eqn:fourier}. (g) Expanded view of the bridge region as a function of time.
 }
 \label{fig:Images}
\end{figure}


The liquid crystal used in these investigations is $8$CB ($4^\prime$-\textit{n}-octyl-$4$-cyanobiphenyl, Sigma-Aldrich), which at room temperature ($22^\circ$C) is in the smectic~A phase, with an in-plane viscosity $\eta=0.052 \unit{Pa\cdot s}$ \cite{schneider_measurement_2006} and a layer thickness $d=3.17\unit{nm}$ \cite{davidov_high-resolution_1979}. The surrounding air is assumed to have a viscosity $\eta ' = 1.827\e{-5}\unit{Pa\cdot s}$ \cite{weast_handbook_1973}. We drew freely-suspended films with $N_b$ in the range $2$--$8$ smectic layers across a $3 \unit{mm}$-diameter, horizontal opening in a glass cover slip and generated islands with radii in the range $4$--$100 \unit{\mu m}$ and thickness $N_i \alt 30$ layers. The large difference in thickness $\Delta N$ between the islands and the background film  gives good optical contrast and helps stabilize the island size for up to an hour against draining, a slow process in $8$CB films in which material is transported from the film into the meniscus at the film boundary. The line tension of an elementary layer step $\lambda_1=\lambda/\Delta N$ was found previously for $8$CB to be $10.5 \pm 0.6\unit{pN}$ by Zou et~al.~\cite{zou_line_2007}, $7 \pm 1 \unit{pN}$ by G\'{e}minard et~al.~\cite{geminard_meniscus_1997}, and $10 \pm 1 \unit{pN}$ in our measurements of the boundary deformation of sessile islands in a vertical film pulled by gravity to the meniscus at the bottom of the film.

The films were enclosed in a sealed chamber in order to minimize disturbances from the surrounding air. Pairs of islands were selected and then brought close together using optical tweezers, eliminating having to wait for the islands to approach each other by random diffusion. In $8$CB films, even when two islands touch, they often do not coalesce immediately but once they began to merge, we recorded the shape evolution of the merged island at rates of up to $67{,}065$ frames/sec (fps)  using a high-speed digital video camera (Phantom v12.1, Vision Research) with $512 \times 512$ pixels, providing a spatial resolution of about $0.5 \unit{\mu m}$.

\section{Island coalescence}

A typical coalescence event (two $27$-layer islands of approximately the same size merging in a $4$-layer film, captured at $21{,}005 \unit{fps}$) is shown in Fig.~\ref{fig:Images}. Soon after the islands connect (at $t=0$), a narrow, fluid bridge forms between them. The bridge then expands and the initially sharp, concave cusps on either side of the bridge move steadily apart and become rounded. Once the bridge spans the width of the merged island, its ends become convex  (at $t\approx 5 \unit{ms}$), following which the line tension slowly forces the merged island to change from approximately elliptical to circular. During the entire coalescence process, the two islands remain along a fixed symmetry axis passing through their centers. The bridge expands normal to this axis and has a well-defined boundary, allowing us to measure the bridge width accurately to within $\pm 1\unit{\mu m}$ for $t \agt 0.5 \unit{ms}$.
We have analyzed the island coalescence in order to determine the bridge width vs.\ time first by measuring directly the neck width of thresholded coalescence images, and second by analyzing time-series images (kymographs) created by making intensity scans parallel to the bridge (see Appendix B for a more detailed explanation of this procedure). These analysis methods produce similar results.
At early times, leading up to coalescence, the island boundaries near the point of contact are very close together and essentially tangent to each other. Since the island boundaries are extremely narrow, with widths that are below the optical diffraction limit, they cannot be resolved individually in the microscope, giving an apparent bridge even before coalescence begins. The initiation of coalescence of the two islands (where $w=0$) was therefore taken to be the instant at which the perceived bridge width began to grow.

\begin{figure}
 \centering
 \includegraphics[width=2.5in]{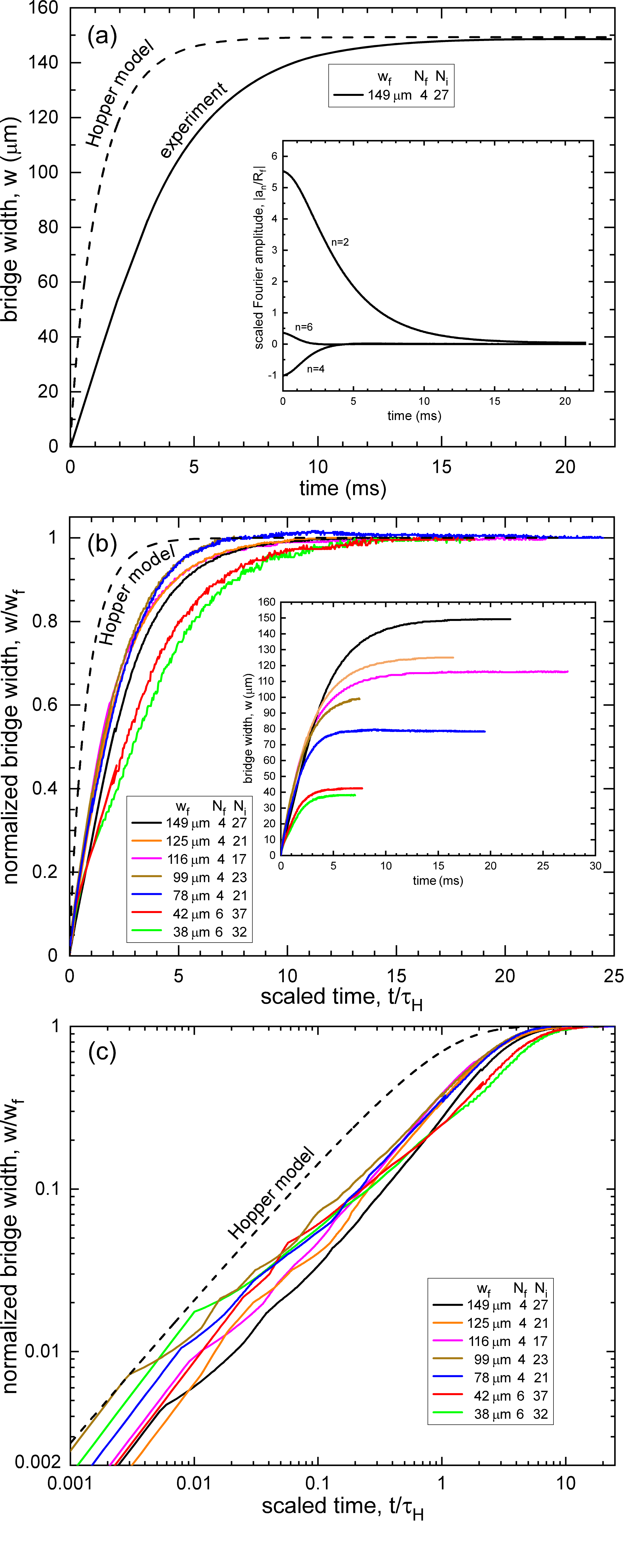}
 \caption{
Bridge expansion during island coalescence.
(a) Bridge width evolution during the coalescence event shown
in Fig.~\ref{fig:Images} ($w_f = 149 \unit{\mu m}$, $N_b=4$, $N_i=27$), beginning with rapid expansion of the connected
region and ending with a slow relaxation towards a circular shape (solid curve). Hopper's model  for isotropic fluid cylinders with the same line tension and viscosity as $8$CB (dashed curve), predicts more rapid dynamics ($\tau_H=1.44 \unit{ms}$) than observed in our experiments: fitting the Hopper model to the final stages of evolution yields a characteristic time $\tau=3.4 \unit{ms}$. The effective Saffman length is $\ell_{S}=138 \unit{\mu m}$. The inset shows the evolution of the principal Fourier modes $a_2$, $a_4$, and $a_6$  (scaled by $R_f$) defined in Eq.~\ref{eqn:fourier},  illustrating the decreasing deviation of the boundary from circularity with time.
(b) Normalized bridge expansion observed during the coalescence of several pairs of islands of the same thickness and similar size vs.\ time scaled by $\tau _H$. The legend shows for each coalescence event the final island diameter and the number of layers in the background film and islands.
The bridge expansion scales reasonably well, with the two outliers (the red and green curves) corresponding to thicker background films.
The inset shows the raw bridge expansion data.
(c) Normalized bridge width vs. scaled time on logarithmic axes.
The black experimental curve in all the bridge width plots shows the same coalescence event.
}
 \label{fig:Hopper}
\end{figure}

The boundary of the merged island can be fitted to a Fourier expansion in the first three even powers of $\cos\theta$, using

\begin{equation}
 r(\theta,t)=R_f \bigl (1+\sum a_n(t) \cos(n\theta) \bigr ), n=\{2,4,6\}
 \label{eqn:fourier}
\end{equation}

\noindent
where $R_f$ is the final island radius. The odd modes are absent by symmetry. The Fourier components of order four and six decay relatively rapidly (Fig.~\ref{fig:Hopper}(a) Inset), leaving the second-order term dominant in the final shape relaxation process. The amplitudes of any higher order harmonics are significant only at the earliest stages of coalescence, decaying very quickly once the islands begin to merge.

Even at the fastest observed bridge expansion velocity ($v \sim25 \unit{mm/s}$) and largest relevant length scale ($\ell\sim 150 \unit{\mu m}$) in these experiments, the Reynolds number is small ($\mathrm{Re} =  \rho v \ell / \eta ={\cal O}(0.1)$, where $\rho$ is the density of the liquid crystal). The hydrodynamics in the plane of the film are thus expected to be Stokes-like. The evolution of the bridge width $w$ during the coalescence of a typical pair of islands is plotted in Fig.~\ref{fig:Hopper}(a), together with Hopper's prediction for the coalescence of two infinite, viscous cylinders of isotropic fluid surrounded by inviscid fluid \cite{hopper_coalescence_exact_1984} taking the effective surface tension as
$\gamma = \lambda_1/d=3.15\unit{mN/m}$, where $\lambda_1=10 \unit{pN}$,
and using the viscosity of $8$CB. The coalescence of several pairs of islands of the same thickness and similar size vs.\ time scaled by the Hopper relaxation constant $\tau_H = N_i \eta d R_f/(\Delta N \lambda_1)$   is shown in Fig.~\ref{fig:Hopper}(b), with the raw data plotted in the Inset, and on a log-log scale in Fig.~\ref{fig:Hopper}(c).
In all of the experiments, the growth at early times is observed to be about a factor of two slower than Hopper's model, in accord with the observations of Shuravin et al.\ \cite{Shuravin2019}, suggesting that there are sources of dissipation in addition to those due to material flow in the island interior  arising from processes not present during the coalescence of two isolated fluid disks, such as shear flow in the surrounding film and in the air, that need to be considered.

\section{Dissipation Mechanisms}
\subsection{Flow in the Film, the Islands, and the Air}

\begin{figure}[htb]
 \centering
 \includegraphics[width=3in]{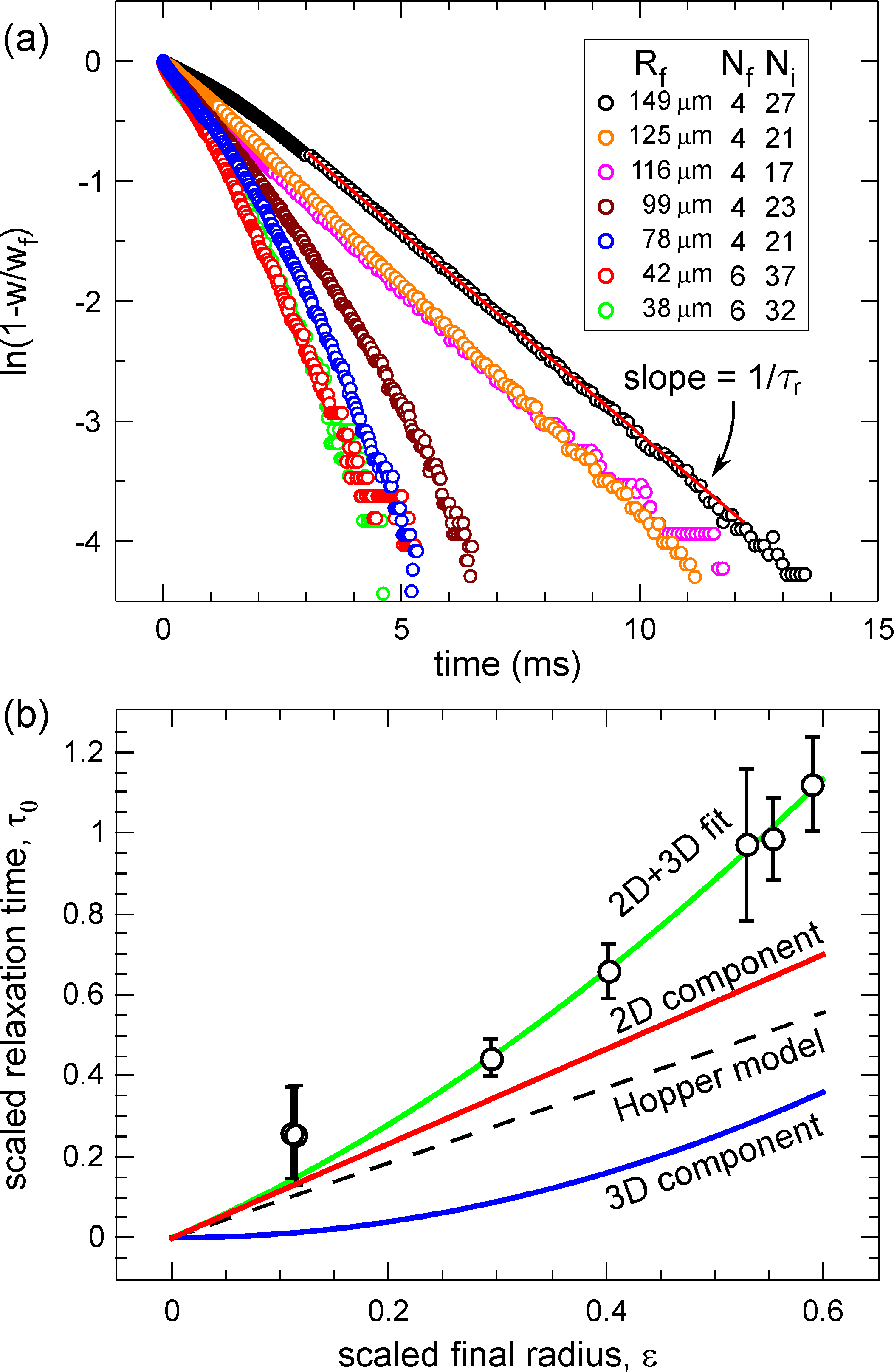}
 \caption{
Merged island shape relaxation at long times. (a) Evolution of the difference between the normalized bridge width $w$  and its final value $w_f$ (the legend shows the corresponding final island radius $R_f$). The response becomes exponential after a few ms, with a slope corresponding here to the inverse time constant, $1/\tau_r$. (b) Scaled relaxation time $\tau_0$ as a function of scaled island radius $\varepsilon$ fitted using Mann's model (Eq.~\ref{eqn:tau0}). In these experiments, the $2$D component (red curve), accounting for dissipation in the background film,  is clearly more important than the $3$D component (blue curve), which describes dissipation occurring in the air, especially for small islands.  The fitted curve (green) gives a value of line tension consistent with previous film experiments.
}
 \label{fig:LineTension}
\end{figure}

During island coalescence, there is significant dissipation associated with the Stokes-like flow in the plane of the film and in the surrounding air.
At long times, the relaxation of the elliptical merged island towards the final circular shape is exponential (Fig.~\ref{fig:LineTension}(a)).  Mann et~al.\ have investigated theoretically the evolution of an analogous system, elliptical lipid domains relaxing in Langmuir films, considering dissipation coming from flow  in the ``surface layer'' (the background film, including the embedded domains) and in the sub-phase (water) \cite{mann_hydrodynamics_1995}.
Adapted to freely-suspended smectic films in air, the Mann model predicts a relaxation time given by

\begin{equation}
\tau_r = \frac{\eta_\mathrm{2D}R_f}{\lambda} + 2\frac{5\pi}{16}\frac{\eta' R_f^2}{\lambda} \; ,
\label{eqn:tau}
\end{equation}

\noindent
where the (now redefined) effective $2$D viscosity $\eta_\mathrm{2D}=(N_i+N_b)\eta d$ combines contributions from dissipation within the island and background films, $R_f=w_f/2$ is the final island radius, and the prefactor of $2$ in the second term is a consequence of the film being in contact with the ``sub-phase''  (air) at both surfaces.  If we scale $R_f$ and $\tau_r$ to be dimensionless, defining $\varepsilon = R_f/\ell_S$, where $\ell_S$ is the Saffman length computed using $\eta_\mathrm{2D}$,  and $\tau_0=\tau_r\Delta N\lambda_1/(\eta_\mathrm{2D}\ell_S)$, then Eq.~\ref{eqn:tau} can be rewritten

\begin{equation}
 \tau_0=\frac{1}{k} \left(\varepsilon+\frac{5\pi}{16}\varepsilon^2 \right) \; ,
\label{eqn:tau0}
\end{equation}


\noindent
where $k=\lambda/(\Delta N \lambda_1)$.  If we allow the line tension per layer $\lambda_1$ to be a free parameter and use Eq.~\ref{eqn:tau0} to fit our data  for islands of different radius (green curve in Fig.~\ref{fig:LineTension}(b)), we obtain a value of $8.5 \pm 0.5 \unit{pN}$, which is in agreement with our measurement of $\lambda_1$ from the boundary deformation of sessile islands
and the value obtained by G\'{e}minard et~al.~\cite{geminard_meniscus_1997} from the evolution of dislocation loop size.
The $2$D term (red curve) resulting from dissipation within the film is seen to be dominant for all final island sizes. The $3$D term (blue curve) resulting from the dissipation in the air is significant when the island radius is large but is negligible at small $\varepsilon$, indicating that the contribution of the air is less important and the dynamics become increasingly two-dimensional in character the smaller the diameter of the merged island is compared to the Saffman length.

In summary, the Mann model shows that the coalescence behavior of the larger islands in thinner films can be described quantitatively and parameter-free by including the dissipation of both the air ($3$D) and outer fluid viscosities ($2$D) in the dynamics.  This suggests that a similar generalization of the Hopper model to include these effects (beyond the scope of this work) would reduce the factor-of-two discrepancy between the characteristic times of the Hopper prediction and our observations.   Finally, we note that the coalescence dynamics of the smallest islands in the thickest background film (red and green curves in Fig.~\ref{fig:Hopper}b) trend as expected from the Mann model, but are anomalously slow.


\subsection{Permeative Flow in the Island Boundaries}

Next, we consider the  dissipation associated  with  displacing  material  from  the  boundary of the  merged  island as it shortens,  a process effectively governed in these experiments by smectic permeation.   Unlike isotropic  fluid drops, where the  surfaces and the bulk can exchange molecules quite freely, the innermost layers of smectic islands are bounded  by  the closed layers of edge dislocation loop structures. While each layer is essentially a two-dimensional, Stokesian fluid, mass transport between layers is limited to permeative flow \cite{degennes1998}. The energetics of coalescence are determined by the line tension of the island boundary, which is proportional to $\Delta h = (N_i - N_b) d$, the difference in thickness between the island and the background film.

An island of initial radius $R_1$ (using Hopper's notation) and total thickness $h_i= N_i d$ comprises a central region of flat layers of area  $A \approx \pi R_1^2$ and excess volume $V_c \approx \pi R_1^2 \Delta h$, and a narrow perimeter region (outlined in red in Fig.~\ref{fig:Sketch}a) of height $\Delta h$, width $\sim h_i/2 \approx \Delta h/2$, length $p_1 = 2\pi R_1$, and an excess volume $V_{p1} \approx 2\pi R_1(\Delta h^2/2) \sim V_c(\Delta h/R_1)$.

When two islands of the same thickness merge, the observed total area appears to be conserved.
However, over the course of the coalescence, the total island perimeter length $p$ decreases from $2 p_1$ to $p_f = 2 p_1/\sqrt{2}$, the volume of the perimeter region being reduced by a corresponding amount $2 V_{p1} (1-1/\sqrt{2})$.
This means that during coalescence, liquid crystal necessarily flows out of the perimeter  region, permeating either into the central, flat region of the island or through the background film and into the film meniscus.  The flow is not optically detectable in the experiments but, in principle, the resulting dissipation further retards the coalescence dynamics.   The constancy of the total island area suggests that during the coalescence, apart from the flow out of the island perimeter,  there is no other substantial exchange  of material  between  the  merging islands and the background film.  This is to be expected since, as Fig.~\ref{fig:Sketch}a shows, the innermost layers of the islands (which determine its area) have fixed volume. There could be a net permeative flow into the islands from the background film, for example if the film area were reduced mechanically \cite{Pieranski1993}, but in the experiments described here there are no such driving forces.

In a coalescence event, surface energy of the shrinking island perimeter is converted into heat in two ways: through viscous drag associated with hydrodynamic flow and by dissipation associated with the movement of molecules out of the perimeter.

We can estimate the relative importance of these two kinds of dissipation, first by estimating how much energy is available in the overall coalescence process. In our experiments, the initial island radii $R_1$  are in the range $15 \unit{\mu m} < R_1 < 50 \unit{\mu m}$ and the island thicknesses in the range $0.05 \unit{\mu m} < h_i < 0.12 \unit{\mu m}$. The net reduction in surface energy in a typical coalescence event is   $\Delta U = 2(N_i-N_b) \lambda_1 (p_1 - p_f) = 2(N_i-N_b) \lambda_1 p_1 (1-1/\sqrt{2}) \sim 2 \times 10^{-14}$~J.

Next we estimate how much energy is dissipated by the permeation of LC molecules out of the perimeter.
The reduction in length of the island boundary generates an internal pressure $P$ within the perimeter region that forces the molecules out, a process that can be understood following  the approach of Oswald and Pieranski \cite{Pieranski1993,Oswald2001,Oswald2003,Oswald2004,Oswald2005} and of Stannarius and co-workers \cite{May2012,May2014,Daehmlow2018,Salili2016,Trittel2019}, who have studied  extensively  the  flow associated  with the motion and evolution  of dislocations  (layer steps)  in smectic  films.
The  permeation flow velocity  is $v_p  = \lambda_p P/\Delta h$, resulting in a flux $dV_p/dt = -v_p p \Delta h$ that is proportional to the total perimeter length $p$. Assuming an $8$CB permeation coefficient $\lambda_p \approx d^2 /\eta \approx 2 \times 10^{-16}  \unit{m^2 /Pa \cdot s}$ \cite{degennes1998},  the pressure in the perimeter region is $P \sim \eta (\Delta h^2/d^2)/\tau$, where $\tau \sim 5 \unit{msec}$ is the total time of the relaxation.  We estimate the resulting net dissipation to be $P\Delta V \sim 3 \times 10^{-18} \unit{J}$. This is about $0.1\%$ of the net total dissipation calculated above, meaning that only a tiny fraction of $\Delta U$ is expended in expelling the molecules from the perimeter, with shear flow in the central region of the island being the dominant loss mechanism during coalescence.

Finally, we consider whether the bridge expansion rate might be limited by permeation. The pressure gradient between the dislocation core and the island boundary, which we assume to be maintained by the line tension, is

\begin{equation*}
    \frac{dP}{dz} \approx \frac{P}{\Delta h/2} \, .
\end{equation*}

\noindent
where $P = \lambda/S$ is the pressure as before and $S = (\Delta h^2/2)$ the cross-sectional area of the dislocation region at the island boundary. If we assume that $\Delta h \approx 100 \unit{nm}$, we have $v_p \approx 130\unit{\mu m/s}$. In principle, the liquid crystal can permeate across the entire outer surface of the dislocation loop, which has an area $p \Delta h$, and flow into the background film.
If the bridge expansion rate were limited only by permeation in the perimeter, and if permeation occurred at the maximum flow rate $dV/dt$, defined above, then the bridge expansion speed would be

\begin{eqnarray*}
    v_{\rm max} = (dp/dt)/2 &=& (dV/dt)/(2S)          \\
                  &=& v_p(p \Delta h)/(\Delta h^2) \\
                  &=& v_p p/\Delta h         \, .
\end{eqnarray*}

\noindent
Assuming a typical boundary length $p \approx 500 \unit{\mu m}$, we obtain $v_{\rm max} \approx  650 \unit{mm/s}$. This is two orders of magnitude larger than the fastest growth velocities observed in our experiments at early times ($\sim 20 \unit{mm/s}$), from which we conclude that the bridge expansion dynamics are not limited by permeation between the smectic layers.

\section{Conclusion}

We have observed directly  the  coalescence of disc-shaped, smectic fluid islands in thin, quasi-two dimensional smectic films in a regime governed mainly by two-dimensional hydrodynamics.  Coalescence of the islands is driven by the line tension along the island boundary, at a rate limited principally by viscous drag associated with shear flow in the island and the background liquid crystal film.
The merged island evolves rapidly from the narrow-waisted shape formed immediately after coalescence to a rounded dumbbell, then becomes elliptical and finally relaxes exponentially to being circular.
Application of Mann's model for the relaxation of fluid domains in Langmuir films, which includes dissipation in both $2$D and $3$D, shows that at long times in the island coalescence process, dissipation occurs principally in two dimensions, in the thin fluid smectic membrane rather than in the surrounding air.

\section{Acknowledgments}
This work was supported by NASA Grants~NAG-NNX07AE48G and NNX-13AQ81G, by the Soft Materials Research Center under NSF MRSEC Grants~DMR~0820579 and DMR-1420736, and by the German Space Management DLR Grant~OASIS-Co (50WM1744).
K.H.\ was also supported by German Science Foundation (DFG) Grants HA8467-1/1 and HA8467/2-1.
We thank R.~Stannarius for fruitful discussions.

\section*{Appendix A: Hopper's Model for the Coalescence of Two Fluid Cylinders}

An exact, analytical solution for the evolution of the shape of two viscous, infinite cylinders of equal size coalescing in an idealized inviscid fluid to form a single cylinder of radius $R_f$ has been derived by Hopper \cite{hopper_coalescence_exact_1984,hopper_plane_1990}. Coalescence is driven in this model by the surface tension $\gamma$ and the dynamics are limited by the viscosity $\eta$.  The flow is assumed to be strictly planar.

The evolution of the island shape at any stage of coalescence can be described  \cite{hopper_coalescence_exact_1984} by a pair of parametric equations
\begin{eqnarray*}
 x(\theta) &=& R_f \left(\frac{1-m^2}{\sqrt{1+m^2} (1+2m\cos 2\theta +m^2)} \right) (1+m)\cos\theta       \\
 y(\theta) &=& R_f \left(\frac{1-m^2}{\sqrt{1+m^2} (1+2m\cos 2\theta +m^2)} \right) (1-m)\sin\theta
\end{eqnarray*}

\noindent
The parameter $m$ varies from $m=1$ (describing the initial state of the system, two contacting circles) to $m = 0$ (describing the final state, a circle of radius $R_f$). The bridge width $w$ is given by
\begin{equation}
 w=2 x(\theta=0)= 2R_f \frac{1-m}{\sqrt{1+m^2}}
 \end{equation}
\noindent
and the time dependence of the bridge width  by
 \begin{eqnarray}
    t        & = & \frac{\pi}{4}\frac{\eta R_f}{\gamma} \int_{m^2}^{1} \frac{1}{\left(\mu \sqrt{1+\mu} \right)   K_1(\mu)} \, d\mu \;  ,
\end{eqnarray}
\noindent
where $K_1(\mu)$ is the complete elliptic integral of the first kind:
 \begin{eqnarray}
    K_1(\mu) & = & \int_0^1\frac{1}{\sqrt{(1-x^2)(1-\mu x^2)}}  \, dx                                                       \\
             & = & \int_0^{\pi/2}\frac{1}{\sqrt{1-\mu\sin^2\theta}} \, d \theta
\end{eqnarray}
\noindent
At long times, the location of the time-dependent island boundary can be  described approximately  in plane-polar coordinates (see Eq.~($64$) in \cite{hopper_plane_1990}) by
\begin{eqnarray}
  R(\theta ,t) \approx R_f \left[1-\exp^{-t/\tau_H} \cos 2\theta \right]
\end{eqnarray}

\noindent
where the characteristic Hopper relaxation time is
$$\tau_H=\frac{\eta R_f}{\gamma} $$

\noindent
For the island coalescence event shown in Fig.~\ref{fig:Hopper}(a) (island layer number $N_i=27$, background film layer number $N_b=4$, layer thickness $d=3.17 \; \rm{nm}$, $8$CB viscosity $\eta=0.052  \unit{Pa \cdot s}$,  final island radius $R_f=74.5 \; \mu$m, and line tension $\lambda_1= 10 \; {\rm pN}$), the equivalent Hopper time is given by
$$\tau_H=\frac{N_id\eta R_f}{(N_i-N_b)\lambda_1}=1.44 \;  \mathrm{ms}$$

Finally, we note that the definition of the complete elliptic integral of the first kind in Hopper's $1990$ paper \cite{hopper_plane_1990} appears to be incorrect, being inconsistent with the rest of that paper and with Hopper's $1984$ publication \cite{hopper_coalescence_exact_1984}: the upper integration bound should presumably be $\pi/2$.

\section*{Appendix B: Experimental Measurement of Bridge Width}

\begin{figure}
 \centering
 \includegraphics[width=3in]{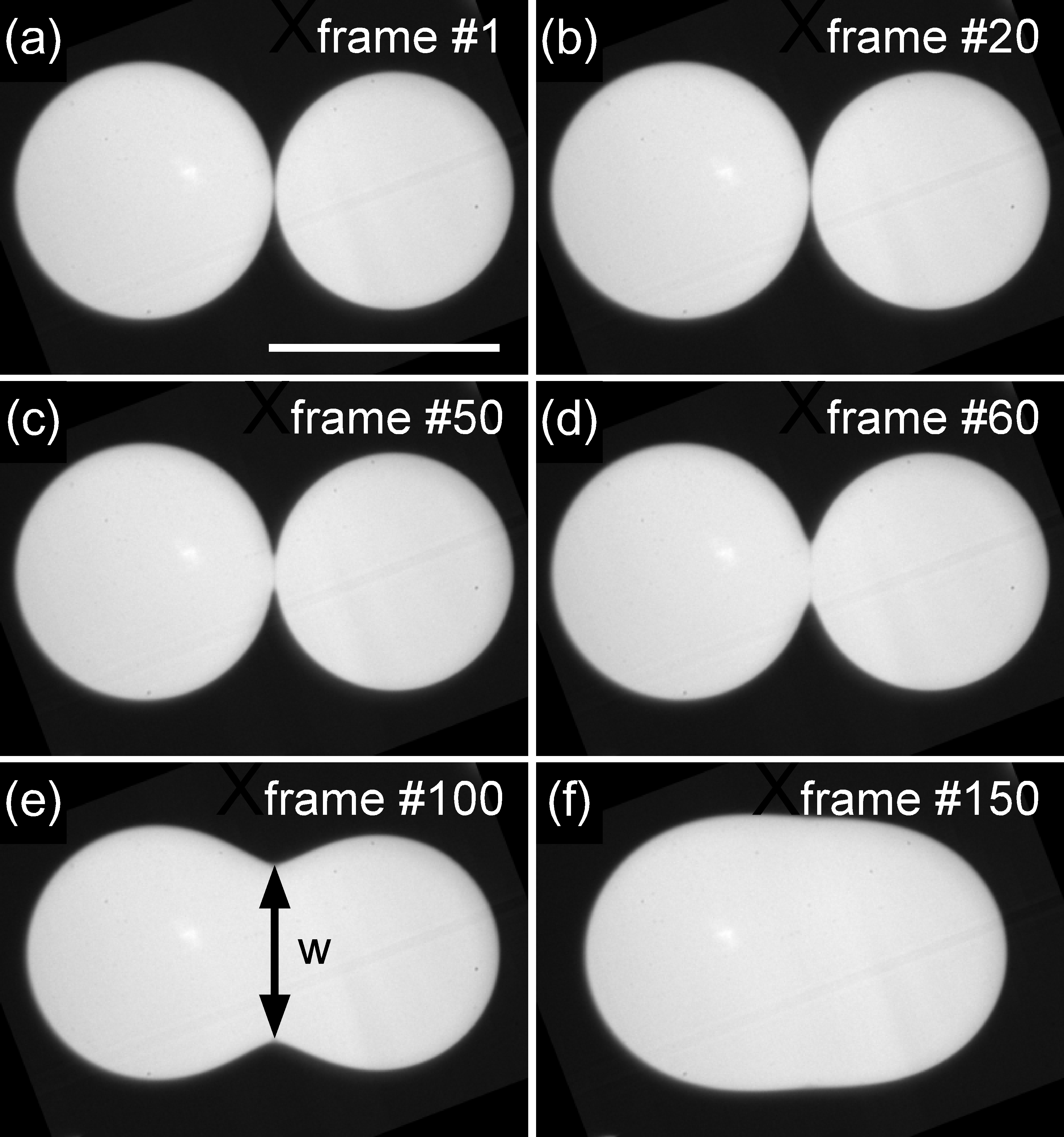}
 \caption{
Coalescence of two $27$-layer thick smectic islands in a $4$-layer film of $8$CB. The scale bar in the first image corresponds to $100 \,\mu$m. The islands in the first two images are in close proximity but have not merged, with coalescence beginning at about frame $40$. The bridge joining the islands is bounded initially by sharp cusps in the merged island boundary but at later times these become more rounded, with the merged island becoming first dumbbell-like, then elliptical, and eventually circular. The images were recorded at $21{,}005 \unit{fps}$.
}
 \label{fig:merging}
\end{figure}

\begin{figure}
 \centering
 \includegraphics[width=3in]{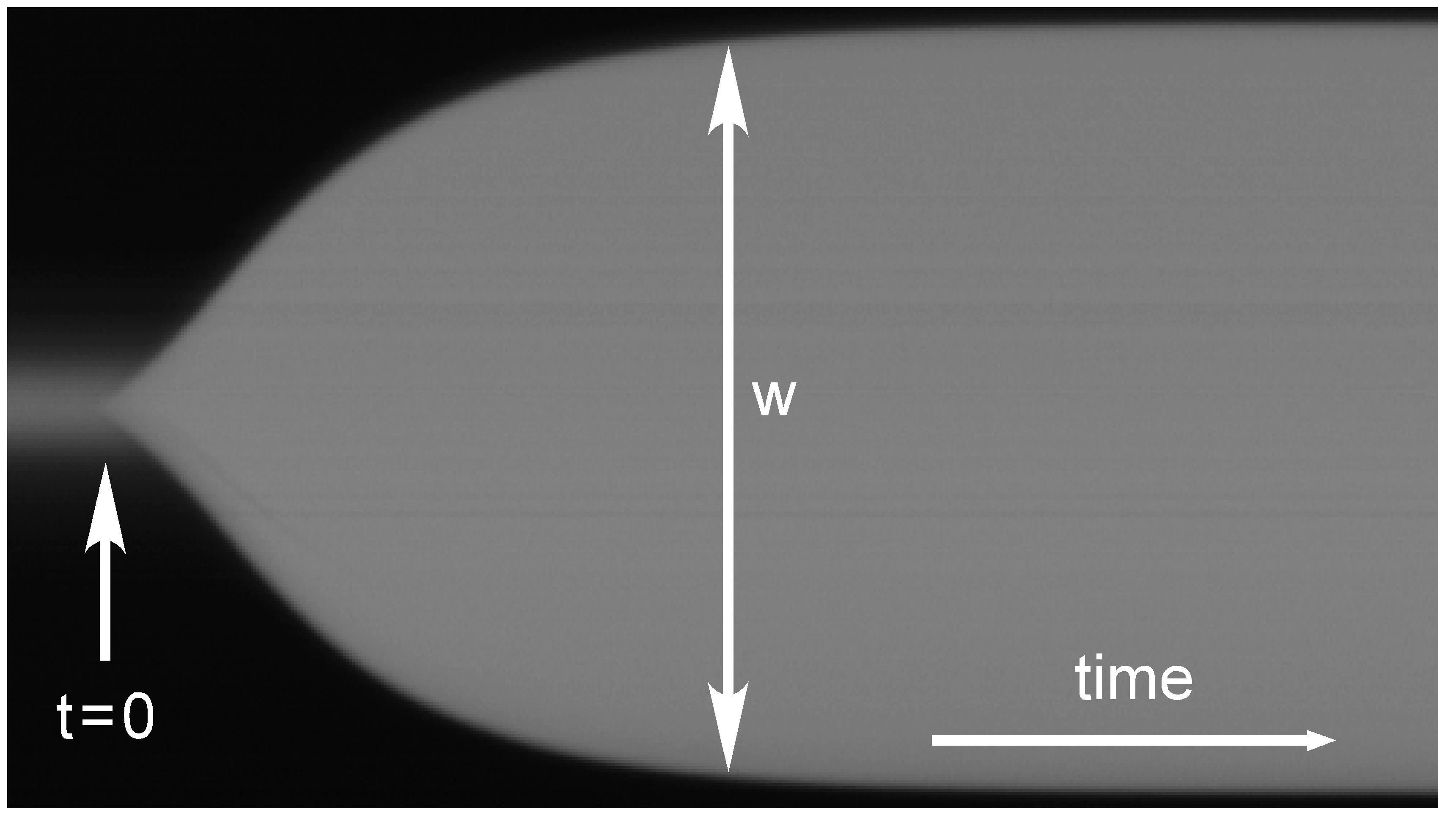}
 \caption{
Kymograph of intensity scans across the coalescence bridge, illustrating the growth in the bridge width, $w$, as a function of time for the event illustrated in Fig.~\ref{fig:merging}. Adaptive thresholding was used to find the bridge boundary locations at early and late times in the coalescence process using, respectively, horizontal and vertical intensity scans of the image. The vertical dimension here is $159 \, \mu$m and the total elapsed time is $24$~ms.
}
 \label{fig:kymograph}
\end{figure}

Selected images from a typical coalescence experiment are shown in Fig.~\ref{fig:merging}. Before the islands merge (which starts at around frame $40$ in this sequence), the two islands are very close, with a separation that is smaller than the optical resolution of the microscope: the thin background film cannot be discerned between the islands and the interfacial region is somewhat brighter than the background film. The initiation of coalescence ($t=0$) was taken to be the time at which the center of the bridge became as bright as the two islands. In order to determine the coalescence bridge width at early times, we constructed time series images (kymographs) from intensity scans across the bridge region carried out on the entire sequence of video frames. The kymograph obtained for the coalescence event depicted in Fig.~\ref{fig:merging} is shown in Fig.~\ref{fig:kymograph}. The positions of the outer edges of the bridge were identified at early times by making horizontal scans through the kymographs and using dynamic thresholding to determine the edge locations, while at later times during the coalescence process, we analyzed vertical scans. Direct measurements of thresholded versions of the original island images carried out for comparison gave similar results.

\clearpage


\bibliographystyle{apsrev}

\end{document}